\documentstyle[rotate,dina4,12pt,psfig]{article}
\newcommand{\be}{\begin{equation}}
\newcommand{\ee}{\end{equation}}
\newcommand{\bea}{\begin{eqnarray}}
\newcommand{\eea}{\end{eqnarray}}

\begin{document}
\title{%
Dynamical fragmentation of flux tubes in the Friedberg-Lee model%
\thanks{Work supported by BMBF and GSI Darmstadt.}}
\author{%
S. Loh\footnotemark[2],
C. Greiner, U. Mosel and M.H. Thoma\footnotemark[3]\\
Institut f\"ur Theoretische Physik, Universit\"at Giessen\\
D-35392 Giessen, Germany\\%
}
\maketitle
\vspace{-1cm}
\begin{abstract}
We present two novel dynamical features of flux tubes in the
\newline Fried\-berg-Lee model. First the fusion of two (anti-)parallel flux
tubes, where we extract a string-string interaction potential which has a
qualitative similarity to the nucleon-nucleon potential in the Friedberg-Lee
model obtained by Koepf et al. Furthermore we show the dynamical breakup
of flux tubes via $q\bar{q}-$particle production and the disintegration
into mesons. We find, as a shortcoming of the present realization of the
model, that the full dynamical transport approach presented in a previous
publication fails to provide the disintegration mechanism in the
semiclassical limit. Therefore, in addition, we present here a molecular
dynamical approach for the motion of the quarks and show, as a first
application, the space-time development of the quarks and their mean-fields
for Lund-type string fragmentation processes.
\end{abstract}

\footnotetext[2]{Part of dissertation of Stefan Loh}
\footnotetext[3]{Heisenberg Fellow}

\section{\bf Introduction}

\bigskip
During the collision of heavy nuclei at ultra-relativistic energies
it is believed that the energy density reached may be
high enough for a quark-gluon plasma (QGP) to be formed. Originally such
collisions were assumed to be highly transparent \cite{AnKoMcLe80,Bjo83} and
the two nuclei pass through each other keeping about half of their energies,
though this might happen only at very high energies. On the other hand
some of the individual target and projectile nucleons
obtain a net color charge due to the exchange of gluons
in the reaction zone. As a consequence, color flux tubes connecting
these color charges, are built up as the nuclei recede. If the ions have
sufficient energy, a large number of flux tubes may be produced, that
may coalesce to a giant color flux tube or simply to an ensemble of
individual flux tubes and color ropes (highly charged flux tubes)
\cite{BiNiKno84,BiCz90,SaSchoSchrSchaGr91}.\\\indent
As kinetic energy of the ions is transferred to the color field
between them it is believed that quark-antiquark and gluon pairs are created
from the field via the Schwinger mechanism. If the pair creation rate is
high enough, the parton density produced may be sufficient for a QGP to be
formed. As the QGP expands and cools, hadronization begins. After
hadronization is complete the hadron gas expands until the strong interactions
freeze out and free hadrons are observed.\\\indent
The present status of model descriptions of this scenario is the following:
On the one hand there are the string- and parton-cascade-models
(Fritjof \cite{AnGuNi87}, Venus \cite{We89}, RQMD
\cite{SoStGr89/2}, HIJING \cite{WaGy91}, parton cascade \cite{GeMu92}),
that are based on phenomenological arguments and (or) perturbative QCD.
However, all the nonperturbative mechanisms like string formation and decay
and the final hadronization, are handled purely phenomenologically, i.~e.~by
fragmentation prescriptions for the string breaking from fitting $e^+e^--$data
or by coalescence models \cite{Ge95,ElGe95}. In order to gain more
insight into the latter mechanism, effective models of QCD can be used, like
the MIT-Bag model \cite{DeGrJaJoKi75}, the NJL-model
\cite{AlReWe95,ChrBlKiPoWaMeArGo96} or the various color dielectric models
\cite{Wilbuch,FbLe77,NiPa82,Bi90}. Within these models there are attempts
to describe the production of quark-antiquark pairs in flux tubes
\cite{PaBr91,FlBi93} or the collapse of flux tubes \cite{WiPu95}.
The main topics attacked in these works are the influence of the finite
transverse size of the flux tube on the Schwinger pair production rate,
or neutralization times for the strings via pair production, but not the full
disintegration of flux tubes into physical mesons, which is needed in
a consistent transport description of heavy ion collisions.\\\indent
In this paper we consider flux tubes constructed in the Friedberg-Lee model,
which has had many successes in other applications \cite{Wilbuch}
and is a fully dynamical field theory with confinement. The quark degrees
of freedom are handled via a transport equation derived by Elze and Heinz
\cite{ElHe89} (see also Bl"ttel et al.~\cite{Blaettel}). In previous publications
\cite{KaVeBiMo93,VeBiMo95,LoBiMoTh96}
we have investigated the dynamical confinement properties in
nucleon-nucleon collision in the colorless version of the model as well as
color-excitations in the full model, whereas here the outline is as follows:\\\indent
In section 2 we give a brief introduction to the Friedberg-Lee model, and
show how the parameters of the model can be fixed by static properties of
hadrons and flux tubes. In section 3 we extract a string-string interaction
potential by a dynamical fusion of two (anti-)parallel strings and in
section 4 we present the full transport dynamical approach to the disintegration
of a flux tube into mesons via quark antiquark particle production. After the
discussion of some intrinsic problems connected with the semiclassical limit of
the model, we give the space-time description of a string-fragmentation process
of the Lund-type within a molecular dynamical approach.

\section{\bf The Friedberg-Lee model}

The nontopological soliton model of Friedberg and Lee \cite{FbLe77,FbLe78},
further developed by Goldflam and Wilets \cite{GoWi82,WiBiLuHe85},
has enjoyed considerable interest in recent years, because it provides a
dynamical description of hadrons as solitons of a selfinteracting
mean field. Static solutions of this model have been studied extensively
during the last decade \cite{Wilbuch,WiBiLuHe85,BiBiWi88},
leading to a satisfactory description of hadronic properties. In its
original version the Lagrangian is given by
\be
  {\cal{L}} = {\cal{L}}_q + {\cal{L}}_{q\sigma} + {\cal{L}}_{\sigma} +
            {\cal{L}}_G \; ,
\ee
where the various terms are as follows:
\be
  {\cal{L}}_q = \bar{\Psi} (i\gamma_{\mu}D^{\mu}-m_0)\Psi
\ee
describes the quarks as Dirac particles of current mass $m_0 \approx 10 \,MeV$.
The fermion wave functions have 4 (Dirac) times 3 (color) times n (flavor)
components and the covariant derivative is $D^{\mu} = \partial^{\mu} -
ig_v ({\lambda^a \over 2}) A^{\mu}_a$, where $A^{\mu}_a$ are the gluon fields
that are coupled to the quarks covariantly. The term
\be
  {\cal{L}}_{q\sigma} = \bar{\Psi} g_0\sigma \Psi
\ee
describes the coupling of the quarks to the $\sigma$-field, which
is assumed to mimic the long range and nonabelian effects of multi gluon
exchange. The kinetic and potential part for the $\sigma$ field
\be
  {\cal{L}}_{\sigma} = {1 \over 2}(\partial_{\mu}\sigma)^2 - U(\sigma)
\ee
contains the self interaction potential $U(\sigma)$ that is supposed to be
quartic in the field,
\be
  \label{uself}
  U(\sigma) = {a \over 2!}\sigma^2 + {b \over 3!}\sigma^3 +
              {c \over 4!}\sigma^4 + B \; .
\ee
The constants in (\ref{uself}) are adjusted in such a way that $U(\sigma)$
has a minimum at $\sigma = 0$ and another, energetically lower one,
at the nonzero vacuum expectation value $\sigma_{vac}$, where the potential
is assumed to vanish. The constant $B$ is chosen such that
$U(\sigma_{vac}) = 0$. Since also $U(0) = B$, $B$ can be identified with the
MIT-Bag constant, or the volume energy density of the cavity.\\\indent
Color gluon fields are introduced as in QCD except for an interaction with
the soliton field through a dielectric function $\kappa(\sigma)$,
\be
  {\cal L}_G = - {1 \over 4} \kappa(\sigma)F_{\mu\nu}^aF^{\mu\nu}_a \; ,
\ee
with $\kappa(0) = 1$ and $\kappa({\sigma}_{vac}) = 0$. $\kappa(\sigma)$ is
not uniquely defined in this model, and a choice must be made as to its
functional form. Several suggestions have been made in the past, that may
be summarized to the general form
\be
  \kappa_{nm}(\sigma) = | 1 - ({\sigma \over \sigma_{vac}})^n |^m
  \Theta(\sigma_{vac}-\sigma) \; .
\ee
Friedberg and Lee \cite{FbLe77} originally proposed $n=m=1$, while others
\cite{Le79,BiBiMaWi85} have
suggested $n=1$, $m=2$ or even $(n,m)=(2,1)$, $(2,3/2)$
and $(2,2)$ \cite{GoWi84}. We prefer the parameters $(n,m) = (1,2$), for
this choice guarantees a vanishing derivative of $\kappa$ at $\sigma_{vac}$.\\\indent
Color confinement is obtained by the general properties of $\kappa$;
it can be shown that a medium with vanishing vacuum value of the
dielectric constant is color confining, i.e. net color charges would lead to
infinite energy configurations \cite{Wilbuch}.\\\indent
Since $\kappa$ and the $\sigma$-field are supposed to represent the nonperturbative
structure of QCD, the gluon fields $F_{\mu\nu}^a$ are treated as Abelian
(one gluon exchange) Maxwell fields, which obey
\bea
  \partial^{\mu}(\kappa F_{\mu\nu}^a) = j_{\nu}^a  \; , \\
  \mbox{with} \;\;\; F_{\mu\nu}^a = \partial_{\mu}A_{\nu}^a -
  \partial_{\nu}A_{\mu}^a \; ,
\eea
with the color current density
\be
  j_{\nu}^a = -ig_v\bar{\Psi}\gamma_{\nu}{\lambda^a \over 2}\Psi \; .
\ee
All the non-Abelian features of QCD are assumed to be contained in the
dielectric function $\kappa(\sigma)$. The other two field equations derived
from the Lagrangian are then
\be
  \label{dirac}
  (\gamma^{\mu}(i\partial_{\mu}-ig_v{\lambda_a \over 2}A_{\mu}^a)
  -m_0-g_0\sigma)\Psi = 0
\ee
for the quarks and
\be
  \label{klegor}
  \partial_{\mu}\partial^{\mu}\sigma + U'(\sigma) + {1 \over 4}\kappa'(\sigma)
  F_{\mu\nu}^aF^{\mu\nu}_a + g_0\rho_s = 0
\ee
for the $\sigma-$field, with the scalar density $\rho_s = \bar{\Psi}\Psi$. The
primes on $U'$ and on $\kappa'$ denote differentiation with respect to
$\sigma$.\\\indent
We use the following approximations: The $\sigma$ and the $A_{\mu}^a$-field
are treated as classical fields (mean field approximation). For the gluon field
we choose the Coulomb gauge $\vec{\nabla}\cdot(\kappa\vec{A}^a) = 0$ resulting in
\bea
  \label{maxsca}
  \vec{\nabla}(\kappa\vec{\nabla}A_0^a) &=& -j_0^a \; , \\
  \label{maxvec}
  -\kappa\partial_t^2 \vec{A}^a + \vec{\nabla}^2\kappa \vec{A}^a -
  \vec{\nabla}\times (\kappa \vec{A}^a \times {\vec{\nabla}\kappa \over \kappa})
  &=& - \vec{j} + \kappa\vec{\nabla}\partial_tA_0^a \; .
\eea
In the following equation (\ref{maxvec}) is neglected as it can be shown, that
currents within a flux tube do not produce a magnetic field because the
displacement current is exactly cancelled by the convection current
if the string radius stays nearly constant \cite{WiPu95}. As a consequence,
we determine the colorelectric field instantaneously by
$\vec{E}^a = -\vec{\nabla}A_0^a$, which is exact in pure 1+1-dimensional
electrodynamics.\\ \indent
The densities needed as sources for the equations (\ref{klegor}) and
(\ref{maxsca}) are provided by a transport equation derived by Elze and Heinz
\cite{ElHe89} in the semiclassical limit of an exact equation of motion for
the quantum Wigner function making use of the Dirac equation (\ref{dirac}). The
corresponding equations for the phase space distribution functions for quarks
($f$) and antiquarks ($\bar{f}$) read \cite{LoBiMoTh96}:
\bea
  \label{vlas1}
 & & (p_{\mu}\partial^{\mu}-m^*(\partial_{\mu}m^*)\partial^{\mu}_p)f(x,p) =
  g_vp_{\mu}F^{\mu\nu}\partial_{\nu}^pf(x,p)  \\
  \label{vlas2}
 & & (p_{\mu}\partial^{\mu}-m^*(\partial_{\mu}m^*)\partial^{\mu}_p)\bar{f}(x,p) =
  -g_vp_{\mu}F^{\mu\nu}\partial_{\nu}^p\bar{f}(x,p) \; .
\eea
We define the effective mass $m^* = m_0 + g_0\sigma$, the energy
$\omega=\sqrt{\vec{p}^2+m^{*2}}$ and a factor $\eta = 4$ describing spin and
isospin degeneracy. In terms of $f$ and $\bar{f}$ the scalar density is then
given by
\be
  \label{dens}
  \rho_s = {\eta \over (2\pi)^3} \int d^3p {m^* \over \omega}
           (f(x,p)+\bar{f}(x,p)) \; ,
\ee
and the color charge density by
\be
  \label{jnull}
  \rho = j_0 = {\eta \over (2\pi)^3} \int d^3p (f(x,p)-\bar{f}(x,p)) \; .
\ee
In deriving (\ref{vlas1}) and (\ref{vlas2}) the Abelian limit of the model
has been used to reduce the color octet components of the currents and fields
to one component denoted by $j_0$ and $A_{\mu}$. Since all the nonabelian
effects of QCD are assumed to be modeled by the $\sigma-$field, this assumption
is justified. We therefore drop the color index here and in the following.\\\indent
The equations (\ref{vlas1}) and (\ref{vlas2}) are a set of usual Vlasov
equations describing the motion of charged particles in a selfconsistently
generated scalar and vector field, determined by equation (\ref{klegor}) and
(\ref{maxsca}), respectively. They are solved by the so called {\em testparticle
method} developed by Wong et al.~ \cite{Wo82}, where every physical
(anti-)quark is represented by an ensemble of $N_T$ testparticles
\bea
  \label{ensemble1}
  f({\vec{x}},{\vec{p}},t) &=& \sum_{i=1}^{N_T}\delta({\vec{x}}-{\vec{x}}_i(t))
  \delta({\vec{p}}-{\vec{p}}_i(t))
  \; , \\
  \label{ensemble2}
  \bar{f}({\vec{x}},{\vec{p}},t) &=& \sum_{i=1}^{N_T}\delta({\vec{x}}-\bar{{\vec{x}}}_i(t))
  \delta({\vec{p}}-\bar{{\vec{p}}}_i(t))  \; ,
\eea
with the testparticles moving according to the Hamiltonian equations of motion:
\bea
  \label{ham1}
  {d {\vec{x}}_i \over dt} &=& {{\vec{p}}_i \over \omega_i} \; , \\
  \label{ham2}
  {d {\vec{p}}_i \over dt} &=& -{m^*({\vec{x}}_i) \over \omega_i}
  \nabla_xm^*({\vec{x}}_i)+g_v{\vec{E}}({\vec{x}}_i)  \; .
\eea
The testparticle coordinates for the antiparticle ensemble obey the same
equations of motion except for $g_v \to -g_v$. The simulations are run in the
so-called full ensemble method.\\\indent
Since we work in the abelian limit of the theory, the color charge of the
quarks can only be positive or negative. While this presents no problems for
the mesons, color neutral baryons, which consist of three quarks, cannot be
constructed in a straightforward way. We therefore work in the latter case
with diquarks, which carry baryon number $B=+2/3$ and negative color charge.
Thus, in the case of mesons, $\bar{f}$ denotes the phase-space distribution
of antiquarks with $B=-1/3$ and negative color charge, whereas in the case
of baryons $\bar{f}$ denotes the diquark distribution, again with negative
color charge, but $B=+2/3$.
%
%
\subsection{Static limit}

Let us first examine the static limit of the transport equations. From quantum
mechanics we know that the color charge density in a hadron has to vanish
locally \cite{moselbuch}
\be
  \label{vanrho}
  j_0(x) = <N|\hat{Q}(x)|N> = \int d^3p (f(x,p)-\bar{f}(x,p)) = 0 \; ,
\ee
because the nucleon ground state $|N>$ is a color singlet and $\hat{Q}$
an (arbitrary) color octet operator. From that we conclude that 
there is no colorelectric field in the groundstate, thus $f = \bar{f}$. As
shown by Vetter et al.~ \cite{VeBiMo95}, the distribution functions have to be
of a local Thomas Fermi type
\be
  \label{thofer}
  f(x,p) = \bar{f}(x,p) = \Theta (\mu-\omega) \; ,
\ee
where we have introduced the Fermi energy $\mu$ for the quarks. After inserting
the local Thomas Fermi distributions (\ref{thofer}) into the integral expression
(\ref{dens}) the fermions can be integrated out to give
\be
  \label{rhos}
  \rho_s = \bar{\Psi}\Psi = {2\eta m^* \over \pi^2} \left[ \mu p_f +
  (m^*)^2\log({m^* \over \mu + p_f}) \right] \Theta(\mu - m^*) \; ,
\ee
for the scalar density and
\be
  \label{rhob}
  \rho_v = \Psi^{\dagger}\Psi = {4\eta \over 3\pi^2} p_f^3 \Theta(\mu - m^*)
\ee
for the quark density. Thus we are left with determining the soliton solution
of the remaining $\sigma$-field equation (\ref{klegor}). The equations
(\ref{rhos}, \ref{rhob}) differ from (28) in \cite{VeBiMo95} by a factor of 2
because of the antiparticle contribution.\\\indent
The soliton solution is con\-struc\-ted using a shoot\-ing me\-thod of
Van Wijn\-gaar\-den-Dekker-Brent \cite{Numerical}. We do not
intend to investigate the total parameter dependence of the properties
of the soliton solutions, since these have been extensively discussed
elsewhere \cite{Wilbuch,VeBiMo95}. Instead we restrict ourselves to adjust the
model parameters in a way to reproduce the quark-number, mean mass of delta
and nucleon and the rms radius of the nucleon for a typical baryon. With the
parameters of this best fit kept fixed, we use the Fermi energy $\mu$ in order
to go from the baryonic solution ($N_Q=\int d^3rd^3p(f(x,p)+\bar{f}(x,p))=3$),
resembling a quark-diquark configuration to the mesonic solution ($N_Q=2$),
resembling a quark-antiquark configuration. The parameter set used to obtain
these solutions is displayed in table \ref{tab1}, where we used
$\mu = 1.768\, fm^{-1}$ for the baryon and $\mu = 1.9582\, fm^{-1}$ for the meson.
The physical properties of these soliton
solutions are shown in table \ref{tab2} and \ref{tab3}, where the meson mass,
which typically is too high in these models, can be significantly reduced by
the inclusion of momentum projection methods and the colormagnetic energies
\cite{Wilbuch}. The remaining properties are very well reproduced within
our model approach. These groundstate solutions will be used in the following
sections for dynamical applications.
\section{The string-string interaction potential}
Gluonic interactions in soliton bag models have been considered in two ways
in the literature.
On one hand the spin dependent one gluon exchange (OGE) interaction has
been used to fit the strong coupling constant of the model to the
experimentally observed $N-\Delta-$splitting. In these calculations a gluonic
propagator developed by Bickeboeller et al. \cite{BiBiMaWi85} is used,
resulting in a value for $\alpha_s \approx 2$
\cite{BiBiWi88,HaLi85,DoWi88,AoHy89,Sa93}. Although it is in principle
possible to determine the OGE matrix elements for dynamical systems
without any special spatial symmetry, the calculation of the propagators
is very involved and has therefore not yet been performed.
On the other hand there have been flux tube solutions constructed with an
infinite (cylindrical) geometry assuming a large mass for the quark-antiquark
pair at the endcaps of the tube (Born-Oppenheimer approximation) and an
instantaneous colorelectric field between them \cite{BiBiMaWi85,GrGy91}.
Within these approximations the most im\-por\-tant property of the tube, the
string tension $\tau \approx 1\, GeV/fm$, has been fitted, where again values
of the strong coupling constant $\alpha_s \approx 2$ have been ob\-tained.\\\indent
We proceed differently here and calculate first the response of the colorelectric
field to an adiabatic separation of the quark-antiquark pair of the meson
groundstate, i.~e.~the quark and antiquark move apart with a given constant
velocity which is taken to be $v_{q} = -v_{\bar{q}} = 0.1c$ in order to
guarantee adiabacity in our calculations. The quark degrees of freedom are
taken as frozen, so that the spatial shape of the quark distributions is not
changed during this process.
The corresponding field equations (\ref{klegor}) and (\ref{maxsca}) are solved
selfconsistently throughout the time evolution via a staggered leapfrog algorithm
for the complete (3-dimensional) $\sigma$-field \cite{VeBiMo95} combined with
a (2-dimensional) finite element method for the colorelectric field
\cite{mitchell}, exploiting the cylindrical symmetry of the system.
In this way we dynamically generate a color dielectric flux tube with a length
of $8\, fm$ and a width of $\approx 1\, fm$ as described
in \cite{LoBiMoTh96} and fix the strong coupling constant $\alpha_s$ to a
string constant of $\tau = 1\, GeV/fm$, obtaining $\alpha_s = 1.92$, in
agreement with the earlier estimates discussed above.\\\indent
Similarly we can construct a two flux tube configuration with either parallel
or antiparallel colorelectric fields as schematically depicted in figure
\ref{schematic}. Starting with this configuration we let the quarks inside the
different tubes move with a constant and fixed fusion velocity $v_f = 0.1\, c$
towards each other keeping their spatial shape fixed througout the fusion
process. We have checked that at this low fusion velocity the $\sigma-$field
follows the quark motion without retardation. We finally extract an interaction
potential $V_{ad}$ for the two strings via
\cite{KoWiPeSta94}
\be
  V_{ad}(\rho) = E_{\sigma}(\rho) + E_{gl}(\rho)
                 - E_{\sigma}(\infty) - E_{gl}(\infty) \; ,
\ee
with
\bea
  E_{\sigma} &=& \int d^3r \left( {1 \over 2} (\nabla \sigma)^2 + U(\sigma)
               + g_0 \sigma\rho_s(\sigma)\right) \; , \\
  E_{gl} &=& {1 \over 2} \int d^3r \kappa(\sigma) \vec{E}^2 \; .
\eea
The interaction potential
is taken to be the difference in the field energies compared to infinitely
separated strings as a function of the transverse distance $\rho_t$ of the
tubes. \\\indent
Due to the loss of the cylindrical symmetry in this scenario, we adopt the
following prescription for determining the colorelectric field:
We take the axis connecting the quark and antiquark of one of the flux tubes
as the symmetry axis. The colorelectric field is then calculated taking only
the color charges within this tube into account, resulting in an 'undisturbed'
field configuration $\vec{E}_u$, as for instance shown in one of the tubes in
the upper picture of figure \ref{strfus}. The total colorelectric field is
assumed to be given by
\be
  \vec{E}(x,y,z) = \vec{E}_u(x-\rho_t,y,z) \pm \vec{E}_u(x+\rho_t,y,z) \; ,
\ee
where the plus sign is valid for parallel tubes and and the minus sign for
antiparallel ones. The index $u$ denotes the 'undisturbed' field configuration
and $\rho_t$ stands for the (transverse) distance of the symmetry axis of the
individual tube to the x-axis. We consider this coherent addition of the
undisturbed colorelectric fields to be a strong assumption, since we neglect
the direct interaction of the individual quark with both quarks of the
neighboring flux tube. However, in the limit $\rho_t \to 0$, the cylindrical
symmetry is restored and our prescription is valid again. Furthermore, the
equation for the $\sigma-$field is solved with the full 3+1-dimensional technique
developed by Vetter et al.~ \cite{VeBiMo95}.
A complete 3-dimensional finite-element treatment of both, the Poisson equation
(\ref{maxsca}) and the $\sigma-$field equation (\ref{klegor}), is in preparation
\cite{Traxler}.\\\indent
The result for the {\em parallel} configuration is shown in figure \ref{strpot1},
where one can see that there is no long range interaction in this model due
to confinement \cite{GrGy91} for distances larger than $1\, fm$.
But as soon as the tubes get in touch with each other at distances less than
$1\, fm$, the confinement wall between them breaks down and we gain surface
energy from the $\sigma-$field (\ref{strpot1}). The short range behaviour is
strongly repulsive, which is clarified in figure \ref{strfus}, where
the field configurations are shown for the $\sigma-$field (left side) and the
colorelectric field (right side) throughout the adiabatic fusion: when the
charges start to overlap, the coherent addition of the colorelectric field leads
to a doubling of the $\vec{E}-$field and thus to 4 times the colorelectric
energy in the complete overlap configuration. This rise in energy results in
the strong short range repulsion of the parallel strings. The adiabatic
string-string potential is qualitatively very similar to the nucleon-nucleon
potential in the same model obtained by Koepf et al.~ \cite{KoWiPeSta94}, which
has been calculated as the energy difference of a deformed
six-quark bag and two noninteracting separated nucleons. We even get quantitative
agreement with those calculations if we scale the potential with the length
$L$ of the flux tube, which is shown in figures \ref{strpot1} and \ref{strpot2}.\\\indent
For the {\em antiparallel} configuration (figure \ref{strpot2})
again there is no long range interaction. However, as soon as
the confinement wall of the flux tubes starts to break down, we get an
attraction coming from the decrease in energy of the $\sigma-$field.
An even larger attraction results from the
colorelectric field, since the two opposite color charges at the
end of the tubes neutralize each other, so that the color field vanishes at
very low transverse distances and the flux tube collapses.\\\indent
Our model then supports the independent string picture, which is,
for instance, phenomenologically used in the Lund code \cite{AnGuNi87}.
That means that the parallel strings repel each other, building an ensemble
of single individual strings. Antiparallel strings, on the other hand, do
prefer to fuse and neutralize the color charges at the endcaps, leading to a
collapse of the flux tube.

\section{Flux tube breaking}

In this section we show how a flux tube of the Friedberg-Lee model breaks up
due to quark-antiquark pair production. One usually describes such processes
with the Schwinger formula \cite{PaBr91}, which gives a constant pair
production rate of electron-positron pairs in QED, depending only on the
absolute value of the electric field. However, this formula
cannot be naively transferred to the QCD case of $q\bar{q}-$production in a
flux tube for very different reasons.\\\indent
First of all, the back reaction of the produced pairs on the external field
has to be considered \cite{GlMa83}, since this screening of the field is finally
responsible for the breakup of the strings. Furthermore one must consider
the modifications of the pair production rate due to transverse and longitudinal
confinement
\cite{SaSchoSchrSchaGr91,PaBr91,FlBi93,SaSchoSchaMuGr90,SaHoSchaGr92,WoWaWu95}.
Besides many other issues, these authors have shown in various bag models, that
the production rate is strongly suppressed due to the reasons mentioned above.\\\indent
However, a detailed analysis of the time dependent pair production process
within the Friedberg-Lee model has not yet been performed. Therefore we have
to rely on different phenomenological arguments, providing us with a simple
guideline to the space-time evolution of a fragmenting string.\\\indent
The earliest
and most successful model of string fragmentation is the Lund-model
\cite{AnGuInSj83}. Within this model one assumes the quarks to be massless
and therefore moving with the speed of light. Even the original $Q\bar{Q}-$pair,
generating the string, is supposed to travel on the lightcone. Within
these assumptions it is sufficient to describe the fragmentation process
by a 1+1-dimensional space-time geometry, for it is causally impossible
that the produced $q\bar{q}-$pairs propagate towards the endcaps (the
$Q\bar{Q}-$pair). \\ \indent
Four of the most simple string fragmentation diagrams emerging
from this model are shown in the figures \ref{lund2} a) - \ref{lund2} d). The
figures show the propagation of the original $Q\bar{Q}-$pair and the
consecutively produced $q\bar{q}-$pairs under the influence of a constant and
therefore confining colorelectric field. In figure \ref{lund2} a) we show the
breakup of the string due to the production of one pair in the center of the
string and the symmetric and equal-time production of two pairs in figure
\ref{lund2} b). In both cases excited meson states emerge, travelling with the
rapidity of the orginal quarks, whereas in the second case an excited
'Yo-Yo'-state with vanishing velocity is formed in addition. In figure
\ref{lund2} c) and \ref{lund2} d) the
breakup of a string via equal- and non-equal-time production of three
$q\bar{q}-$pairs is presented. In the first case the final excited 'Yo-Yo'-modes
again have a vanishing total momentum whereas in the latter case their total momentum
is nonvanishing, with its value depending on the difference in production time.\\\indent
There are two physical properties of the final mesons that are fixed by the
model parameters \cite{AnGuNi87,AnGuInSj83}: the mass of the final fragments
can be fixed by choosing the space-time breakup points appropriately, according
to
\be
  t^2 - x^2 = 2m^2 / \tau^2 \; ,
\ee
where $m$ is the mass of the final meson and $\tau$ the string-constant. The
second property to be fixed is the rapidity distribution of the final fragments,
that can be chosen through the so called {\em fragmentation function}
\cite{AnGuNi87,AnGuInSj83}.
%
%
Within the range of only a few parameters, that are fitted to $e^+e^--$data, a
very reasonable description of particle spectra for hadron-hadron and heavy-ion
collisions can be obtained. Motivated by these results we present in the following
a fully dynamical space-time description of the fragmentation processes shown
in figure \ref{lund2} within our model approach.
\subsection{The Transport Dynamical Model}
As a prerequisite we assume that the original $Q\bar{Q}-$pair travels
with the speed of light along the $\pm z-$axis, irrespective of the interior
dynamics, which is generated by solving selfconsistently the transport equations
(\ref{vlas1}) and (\ref{vlas2}) together with the mean-field equations for
the $\sigma-$field (\ref{klegor}) and the colorelectric field (\ref{maxsca})
with the numerical techniques described in the previous section.
The source densities entering the field equations are determined via the
integral relations (\ref{dens}) and (\ref{jnull}). The produced $q\bar{q}-$pairs
are inserted into the dynamically evolving flux-tube by assuming their groundstate
spatial shape that has been determined in the previous section and with a
vanishing total momentum \cite{WiPu95}. Each quark of the produced $q\bar{q}-$pairs
is represented by a testparticle ensemble with $N_T = 50000$.\\\indent
In figure \ref{lund1dyn} we show the color charge density $\rho = j_0$, the
energy density of the colorelectric field $\epsilon={1 \over 2}\kappa{\bf E}^2$
and the scalar
$\sigma-$field along the longitudinal and transverse axis. In the upper row,
at $t=1.8\, fm/c$, the quarks have a distance of $3.6\, fm$ and the colorelectric
field between them together with the confining cavity of the $\sigma-$field
builds up. At $t=2\, fm/c$ the $q\bar{q}-$pair is inserted and instantaneously
torn apart by the force of the colorelectric field. In the second row, at
$t=2.6\, fm/c$, we see that the separating $q\bar{q}-$pair causes a screening
of the colorelectric field in the region between them, where the energy density
of the color field already vanishes. Consequently, but with a little delay in
time, the $\sigma-$field is beginning to snap off in the
screening region. In the lower two rows, at $t=3.4 \, fm/c$ and $t=7.0\, fm/c$
the free propagation of the final string fragments can be seen, where we
observe some slight late oscillations of the $\sigma-$field due to its inert
behaviour caused by the large glueball mass. Overall, the observed time
development corresponds to the expected space-time behaviour shown in figure
\ref{lund2} a).
In the lower row of figure \ref{lund1dyn} we recognize a small
dispersion of the $q$ and $\bar{q}$ distributions, which is caused by the
selfinteraction of the quarks, leading to a screening of the charges:
a charge fragment at the end of one of the two substrings feels a smaller
colorelectric force due to its interaction with the charge fragments in front
of it.\\ \indent
%
As a further example we show in figure \ref{lund3dyn} the space-time evolution
of the string fragmentation via equal time production of three $q\bar{q}-$pairs,
in analogy to figure \ref{lund2} c). In the upper row at $t=2.8 \, fm/c$,
the string is already extended to a length of $5.6 \, fm$; the
constant colorelectric field along the string is reflected in the constant
field energy between the charges. At $t=3.0\, fm/c$ the three $q\bar{q}-$pairs
are inserted at $z=0\,fm$ and $z=\pm 2.0\, fm$ with a vanishing relative
velocity.\\\indent
In the second row at $t=3.6\, fm/c$ we once
again see the screening effect from the motion of the quarks inside the
flux-tube and consecutively the beginning of the snapping off of the
$\sigma-$field. In the following we observe, as in the previous example, the
formation of the outer two meson pairs propagating on the lightcone. In the
inner region of the flux-tube, the respective quark pairs penetrate each other,
having already acquired a small dispersion. At the time $t = 4.2\, fm/c$, the
colorelectric field is almost completely screened. This abrupt change of the
colorelectric field, being also a source of the $\sigma-$field (see
eq.~(\ref{klegor})), leaves the $\sigma-$field
highly excited. Therefore the $\sigma-$field snaps off and starts to oscillate
around the nonperturbative vacuum value $\sigma_{vac}$. These oscillations
can be followed in the last two timesteps of figure \ref{lund3dyn}, where
$\sigma \approx 0$ in the center of the string at $z=0\, fm$ and
$\sigma \approx \sigma_{vac}$ around. Comparison with the leftmost column in
figure \ref{lund3dyn} shows that the $\sigma-$field, which is
supposed to localize the quarks, cannot prevent a further dispersion of the
color charges and the latter neutralize and dissolve along the z-axis
to a length of about $8\, fm$ at $t=8.4\, fm/c$. As a consequence the scalar
density of the quarks is reduced by almost an order of magnitude.\\ \indent
In this case the $\sigma-$field undergoes a complete phase
transition to the nonperturbative vacuum, since this is energetically more
favorable, which is clarified in figure \ref{ueff3}. There it can be seen,
that the global minimum of the effective selfinteraction potential
\be
  U_{eff}(\sigma) = U(\sigma) + g_0\sigma\rho_s
\ee
is at $\sigma_{vac}$, if the scalar density drops below $1/3$ of the
groundstate density. After returning to $\sigma_{vac}$, the field undergoes
damped oscillations, irrespective of the remaining $q\bar{q}-$fragments.
%
%
\subsection{The Molecular Dynamical Model}
As already mentioned, in the present numerical realization of our model each
testparticle of the respective ensembles interacts not only with the testparticles
of the other quarks, but also with the testparticles of the same quark. This
selfinteraction, which leads to the dispersion of the quark
distributions, becomes dominant in the limit of only a few interacting
physical particles; in the extreme case, a single color charge distribution
with an almost vanishing mass would immediately dissolve due to the
selfinteraction. Unlike in other models, like e.~g.~the Walecka-model, which
also contains an attractive scalar field and a repulsive vector field and which
has been treated successfully \cite{Blaettel}, in the present case the
attractive $\sigma-$field acts only in transverse direction on the produced
quarks. Its longitudinal effect is completely suppressed by the colorelectric
field.
%
%
Thus we are forced to formulate a new dynamical approach,
that conserves the identity of the different quarks and antiquarks and is free
of the selfinteraction problem. \\ \indent
A possible way to overcome the problems mentioned above is to treat
the quarks in a molecular dynamical framework. In this approach each physical
particle (quark or antiquark) is simulated by just one testparticle and not
by an ensemble of testparticles ($N_T=N_{phys}$) in the equations (\ref{ensemble1})
and (\ref{ensemble2}). Due to the fact that point-like particles
cause short range divergences in the field equations, a specific spherically
symmetric distribution is assigned to each individual testparticle. In our case
we choose this distribution to be the meson groundstate distribution, as we
have calculated in the previous section. This provides us with the scalar
density $\rho_s$ and the charge density $\rho$ of each individual (anti-)quark.
Although violating Lorentz covariance, we keep the spatial shape fixed
throughout the time evolution, as a first approach to the reaction
dynamics. A testparticle then describes the motion of the center of its
distribution according to the Hamiltonian equation of motion (\ref{ham1}).
The initial momentum of the produced (anti-)quarks is chosen to vanish
\cite{WiPu95}, and propagated according to the equation (\ref{ham2}) afterwards.
The total scalar density
and charge density entering the field equations (\ref{klegor}) and
(\ref{maxsca}) is given by the sum over all (anti-)quarks.
The field equations, on the other hand, are simulated by
the same techniques as in the full transport dynamical model.\\\indent
In the following we show how we obtain a space-time description of the
diagram of figure \ref{lund2} d), thus presenting a fully dynamical
description of the disintegration of a flux-tube into a multiple (excited)
meson state within the Friedberg-Lee model.\\\indent
The corresponding space-time evolution of this process can be followed in
figures \ref{lund4dyn} and \ref{lund5dyn}, where once again the color charge
density $\rho$, the energy density of the colorelectric field $\epsilon$ and
the $\sigma-$field is presented at eight different time steps, which are chosen
such, that all relevant steps of the diagram \ref{lund2} d) can be followed.
We start our
presentation at $t=2.7\, fm/c$, shortly after the first $q\bar{q}-$pair has
been inserted in the coordinate center. We see that the quarks of this first
$q\bar{q}-$pair already propagate along their respective directions and observe
a prompt screening of the colorelectric field in the center, whereas the more
inert $\sigma-$field has not yet changed significantly. In the second row of
figure \ref{lund4dyn}, at $t=3.2\, fm/c$, the remaining two $q\bar{q}-$pairs have
just been inserted at $z = \pm 2\, fm$ at $t=3.0\, fm/c$ and are moving under
the influence of the colorelectric field.
We see how the screening proceeds not only at the coordinate center, but also
between the other fragments. The confining $\sigma-$field
still shows no reaction to the interior dynamics. In the third row we clearly
see how the two lightcone fragments have been formed as well as the two excited
'Yo-Yo'-modes in between after $5.4 \, fm/c$. The motion of the latter at
this particular timestep is such that the distributions almost completely sit on
top of each other, so that the colorelectric field vanishes.\\\indent
We additionally
clearly see how the confining $\sigma-$field starts to form around the
final fragments. This process is completed after $7.2\, fm/c$ as we find in the
lower row of figure \ref{lund4dyn}, where the $\sigma-$field again undergoes
slightly damped oscillations in the regions of the nonperturbative vacuum.
Due to the different production times of the inner $q\bar{q}-$pairs, the
respective pairs of the final fragments have a nonvanishing total momentum
when forming an excited meson and thus start to propagate along the z-axis as
can also be seen in this figure.\\\indent
The late time evolution of this reaction is
shown in the figure \ref{lund5dyn} at the time steps $t=7.8, 8.6, 10.6, 12.2\,
fm/c$; note the rescaled z-axis for presentational purposes. During this late
time development we find a more or less complete correspondence of the
dynamical behaviour of our model with the 1+1-dimensional Lund-diagram
\ref{lund2} d). We see the propagation of the final fragments as well as the
remaining 'Yo-Yo'-modes of the inner two excited mesons. The latter are
obviously undamped as expected from the simple diagramatic representation.
%
%
\section{Discussion, Summary and Outlook}
We have presented two novel features of flux tubes in the Friedberg-Lee
model. First we have extracted a string-string interaction potential by an
adiabatic fusion of two parallel or antiparallel strings. The potential
is obtained by comparing the colorelectric and the $\sigma-$field energies
of the fused configuration to the (infinitely) separated configuration.
For the parallel strings it shows a qualitative similarity to the
nucleon-nucleon-potential of Koepf et al.~\cite{KoWiPeSta94}, which has been
calculated in a similar way as the energy difference of a deformed six-quark
bag and two noninteracting seperated nucleons. Therefore it seems to be a
general feature of the Friedberg-Lee model, that the interaction of two
color neutral objects (nucleons, parallel strings) can be described in the
following way:\\ \indent
Due to confinement in this model there is no long range interaction (over
distances typically of the order of $1\, fm$ and more). In the intermediate
range, when the confinement wall starts to break down, we have a balance
of gaining surface energy from the $\sigma-$field and increasing the total
energy from the coherent addition of the colorelectric field.
In the short range region the potential is strongly repulsive due to
the unscreened quark color charges at the endcaps in the case of parallel strings,
and in the NN-interaction due to the unscreened (aligned) color spins of the
quarks. In the case of antiparallel strings, however, a strong attraction is
found leading to a neutralization of the color charges and the collapse of
the flux tube.\\\indent
Secondly we have shown how the flux-tube break\-ing pro\-ceeds in the Friedberg-Lee
model. Since no estimates of the pair production rate in this model have been
performed so far, we give a space-time description of some of the simplest
Lund-type diagrams. We have shown the full transport dynamical behaviour of
the quark and antiquark distributions for a string fragmenting via the production
of one $q\bar{q}-$pair and three $q\bar{q}-$pairs at an equal time. In the
first case the space-time behaviour of two meson pairs propagating on the
light-cone can be reproduced by our model. In the latter case of three produced
$q\bar{q}-$pairs, the outer meson pairs are also formed, but in the center
of the fragmenting string, where the two parallel 'Yo-Yo'-modes are supposed
to build up, we observe a different behaviour:\\ \indent
When the respective quark distributions start to penetrate and neutralize,
a small dispersion of these is already observed. At this time the colorelectric
field is completely screened in the center of the string. This sudden
change of the source of the flux-tube leaves the $\sigma-$field highly excited
and thus it starts to oscillate around the nonperturbative vacuum value.
The $\sigma-$field is not able to prevent the further dispersion
of the quark distributions, which neutralize and dissolve along the
string-axis. With the scalar density being significantly reduced in this case,
the $\sigma-$field undergoes a complete phase transition to the nonperturbative
vacuum. \\ \indent
Therefore we have to face the following problems within this realization of
the dynamical model: \\ \indent
First of all the large value of the strong coupling constant $\alpha_s = 1.92$,
being responsible for the large energy density of the colorelectric field,
leads to a very deep flux-tube, with values of the $\sigma-$field up to
$-\sigma_{vac}$, whereas the perturbative region is thought to be around
$\sigma \approx 0$. Therefore the $\sigma-$field is highly excited when the
quark charges are screened, causing large amplitude oscillations that are only
slowly damped. In order to avoid these oscillations and provide a faster
damping, one might think of an additional coupling of the $\sigma-$field to
its chiral partner, the pion-field, in the chirally symmetric O(4)-version of
the model \cite{Ki88,DraBraFae89,NeFiGoAl93}. By this extension, the oscillations
of the $\sigma-$field can be used to excite low mass pion-modes, that disperse
the energy of the $\sigma-$field effectively. \\ \indent
Additionally one could determine a new parameter set of the model. A lowering
of the bag-constant, for instance, would facilitate the string-formation, as
well as the fragmentation. However, the present value of $B=56\, MeV/fm^3$ is
already in the lower region of the ones that are generally used.
A larger value of the scalar coupling constant $g_0$ would facilitate the
formation of bags and thus the localization of the quarks in a perturbative
region.\\ \indent
The second problem is the dispersion of the quark distributions,
which is caused by their selfinteraction. This dispersion is an artefact of
our numerical method, for we deal with only a few quarks that are treated
as classical charge distributions and interacting strongly via a classical
confining potential. In this case the selfinteraction of the color charges
becomes dominant and generates a screening of the charges, which
finally is responsible for the dispersion of these.\\ \indent
We have shown a possible way to overcome these problems by treating the quarks
and antiquarks as point particles with an appropriately chosen charge distribution
as commonly used in molecular dynamical simulations. We have shown within
a first and simple approach, where we keep the spatial shapes of these distributions
fixed, how the Lund-type dynamics of the string-fragmentation can be reproduced
for even more complex Lund-diagrams.
An improvement of the molecular dynamical approach can be achieved by
implementing a time-dependent radius or dipole-moment of the quark
distribution. These multipole moments could account for Lorentz contraction,
polarization effects of the distributions or monopole oscillations.\\ \indent
Summarizing we can say that our model provides us with a useful tool to describe
the full dynamical evolution of string formation and decay via multiple
quark-antiquark production. The aim for future investigations should be to
apply our model to more realistic scenarios, like multiple string production
and decay in relativistic heavy-ion collisions, for describing the formation
of hadrons out of an expanding and cooling quark-gluon plasma.

\newpage
\begin{figure}
\caption{Parallel and antiparallel flux tube configurations drawn schematically.}
\label{schematic}
\end{figure}
%
%
\begin{figure}
\caption{The cavity of the $\sigma-$field and the colorelectric field throughout
the adiabatic fusion process of parallel flux-tubes. The transverse distance
of the tubes is (from top to bottom): $\rho= 1.75, 1.25, 0.75, 0.25\, fm$.
The equidistant contour lines start at $-0.125\, fm^{-1}$ to $0.25\, fm^{-1}$
(in the lower row from $-0.175\, fm^{-1}$ to $0.25\, fm^{-1}$).}
\label{strfus}
\end{figure}
%
%
\begin{figure}
\caption{Adiabatic string-string potential for the parallel configuration.
The solid line shows the sum of the contributions from the $\sigma-$ and
the colorelectric field.}
\label{strpot1}
\end{figure}
%
%
\begin{figure}
\caption{Adiabatic string-string potential for the antiparallel configuration.
The solid line shows the sum of the contributions from the $\sigma-$ and
the colorelectric field.}
\label{strpot2}
\end{figure}
%
%
\begin{figure}
\end{figure}
\begin{figure}
\caption{Space-time representation of some of the simplest Lund
stringfragmentation graphs.}
\label{lund2}
\end{figure}
%
%
\begin{figure}
\caption{The figure shows the time evolution of the color charge density $\rho$,
the energy density of the colorelectric field $\epsilon$ and the $\sigma-$field
for the breakup of the string via the production of one $q\bar{q}-$pair. For
calculating the motion of the quarks the full transport dynamics have been
used. The equidistant contour lines run from $-0.5\, fm^{-3}$ to $0.5\, fm^{-3}$
for $\rho$, from $0.0\, fm^{-4}$ to $4.0\, fm^{-4}$ for $\epsilon$ and from
$-0.15\, fm^{-1}$ to $0.25\, fm^{-1}$ for $\sigma$.
The respective charges of the generating $Q\bar{Q}-$pair are indicated by a
$+$ or $-$ sign.}
\label{lund1dyn}
\end{figure}
%
%
\begin{figure}
\caption{The figure shows the time evolution of the color charge density,
the energy density of the colorelectric field and the $\sigma-$field for
the breakup of the string via the production of three $q\bar{q}-$pairs. For
calculating the motion of the quarks the full transport dynamics has been
used. The equidistant contour lines run from $-0.5\, fm^{-3}$ to $0.5\, fm^{-3}$
for $\rho$, from $0.0\, fm^{-4}$ to $3.0\, fm^{-4}$ for $\epsilon$ and from
$-0.125\, fm^{-1}$ to $0.25\, fm^{-1}$ for $\sigma$.
The respective charges of the generating $Q\bar{Q}-$pair are indicated by a
$+$ or $-$ sign.}
\label{lund3dyn}
\end{figure}
%
%
\begin{figure}
\caption{The effective selfinteraction potential for different values of the
scalar density, starting from the solid line to the dotted line:
$0.110\, fm^{-3}$, $0.072\, fm^{-3}$, $0.050\, fm^{-3}$, $0.043\, fm^{-3}$,
$0.036\, fm^{-3}$, $0.021\, fm^{-3}$, $0.000\, fm^{-3}$.}
\label{ueff3}
\end{figure}
\begin{figure}
\caption{The figure shows the early time evolution of the color charge density,
the energy density of the colorelectric field and the $\sigma-$field for
the breakup of the string via the production of three $q\bar{q}-$pairs at
different time steps $t=2.7, 3.2, 5.4, 7.2\, fm/c$. For calculating the motion
of the quarks the molecular dynamical approach has been used. The equidistant
contour lines run from $-0.35\, fm^{-3}$ to $0.35\, fm^{-3}$
for $\rho$, from $0.0\, fm^{-4}$ to $3.0\, fm^{-4}$ for $\epsilon$ and from
$-0.15\, fm^{-1}$ to $0.25\, fm^{-1}$ for $\sigma$.}
\label{lund4dyn}
\end{figure}
%
%
\begin{figure}
\caption{The figure shows the late time evolution of the color charge density,
the energy density of the colorelectric field and the $\sigma-$field for
the breakup of the string via the production of three $q\bar{q}-$pairs at
different time steps $t=7.8, 8.6, 10.6, 12.2\, fm/c$. For calculating the
motion of the quarks the molecular dynamical approach has been used. The
equidistant contour lines run from $-0.35\, fm^{-3}$ to $0.35\, fm^{-3}$
for $\rho$, from $0.0\, fm^{-4}$ to $3.0\, fm^{-4}$ for $\epsilon$ and from
$-0.15\, fm^{-1}$ to $0.25\, fm^{-1}$ for $\sigma$.}
\label{lund5dyn}
\end{figure}
\clearpage
%
%
\begin{table} [ht]
\center{\begin{tabular}{||c|c||} \hline
parameter set &   \\ \hline
$a$ $[fm^{-2}]$   & 0.0      \\
$b$ $[fm^{-1}]$   & -419.3   \\
$c$ $[1]$         & 4973.0       \\
$B$ $[fm^{-4}]$   & 0.283      \\
$g_0$ $[1]$       & 8.0         \\
$m_0$ $[fm^{-1}]$ & 0.025       \\
$\mu$ $[fm^{-1}]$ & 1.768 / 1.9582 \\
$\eta$            & 4         \\
$\sigma_{vac}$ $[fm^{-1}]$ & 0.253 \\ \hline
glueball mass [$GeV$] & 1.045\\
\hline
\end{tabular}}
\caption{The table shows the parameters used to construct a mesonic and
a baryonic soliton solution of the sigma field equation. The first value of
$\mu$ is used for the baryonic solution, the second one for the mesonic.}
\label{tab1}
\end{table}
\begin{table} [ht]
\center{\begin{tabular}{||c|c|c||} \hline
Parameter set & meson & Experimental Data \\ \hline
$E$ $[MeV]$        & 798        &  465  \\
$RMS$ $[fm]$       & 0.653      &  0.66 \\
\hline
\end{tabular}}
\caption{The table shows the results
of the fits for the meson compared to the experimental data.
The experimental data are the mean values of the pion and the rho-meson
(taken from \protect\cite{Wilbuch}).}
\label{tab2}
\end{table}
\begin{table} [ht]
\center{\begin{tabular}{||c|c|c||} \hline
Parameter set & baryon & Experimental Data \\ \hline
$E$ $[MeV]$        & 1099       &  1087 \\
$RMS$ $[fm]$       & 0.693      &  0.83 \\
\hline
\end{tabular}}
\caption{The table shows the results of the fit for the baryon compared to the
experimental data. The experimental data are taken from \protect\cite{Wilbuch}.}
\label{tab3}
\end{table}
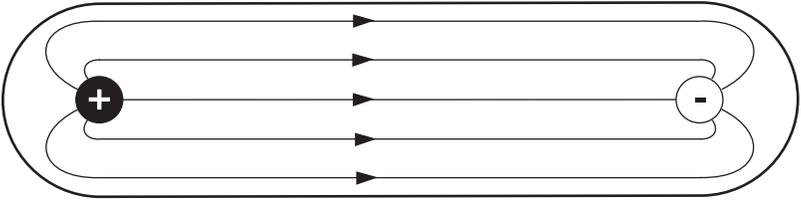
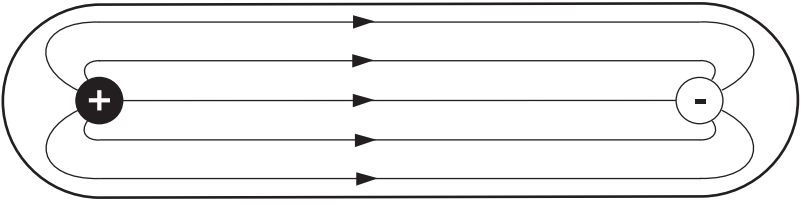
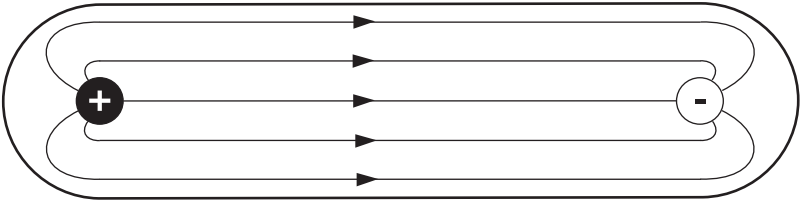
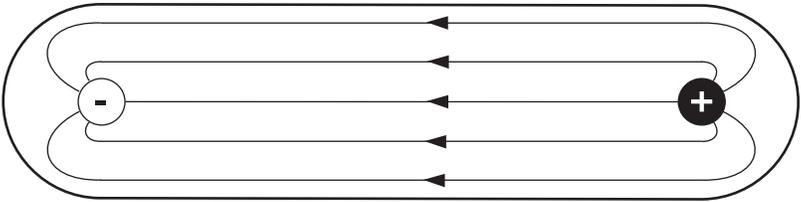
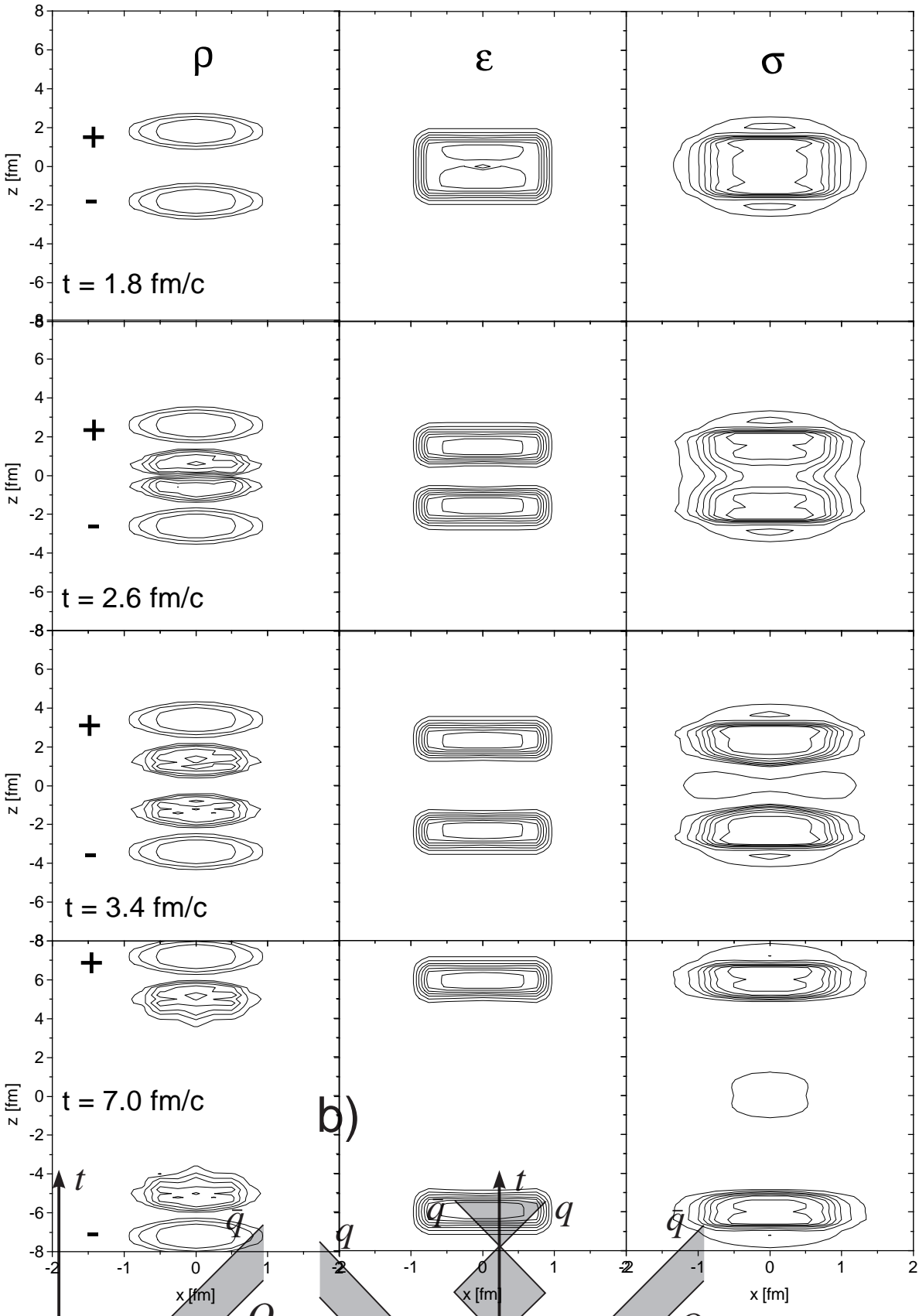
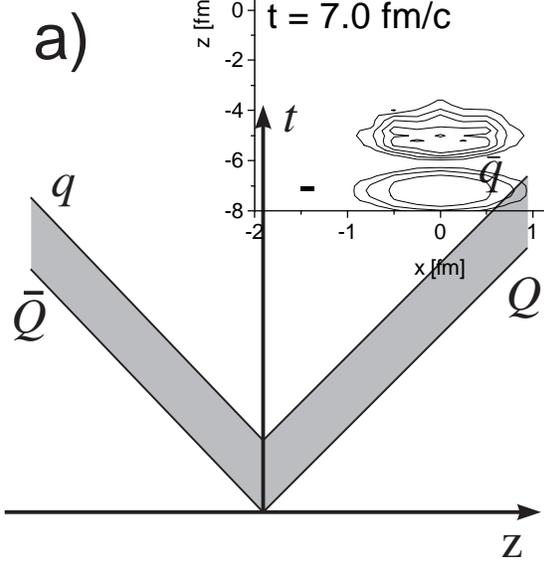
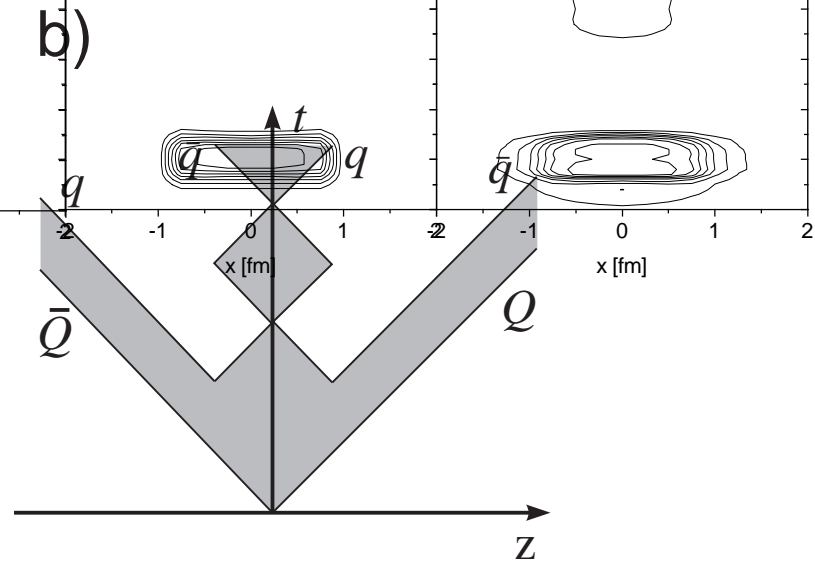
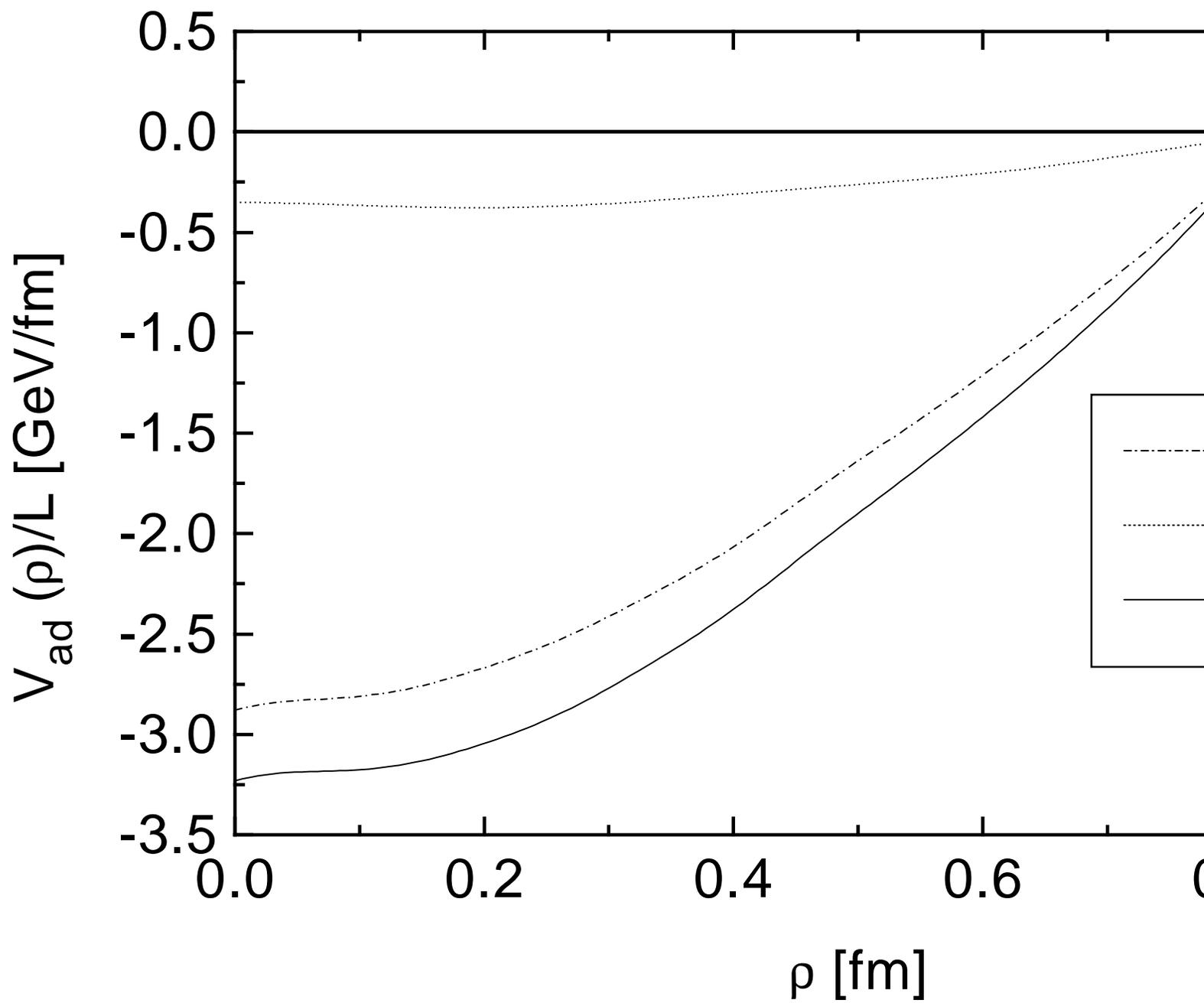
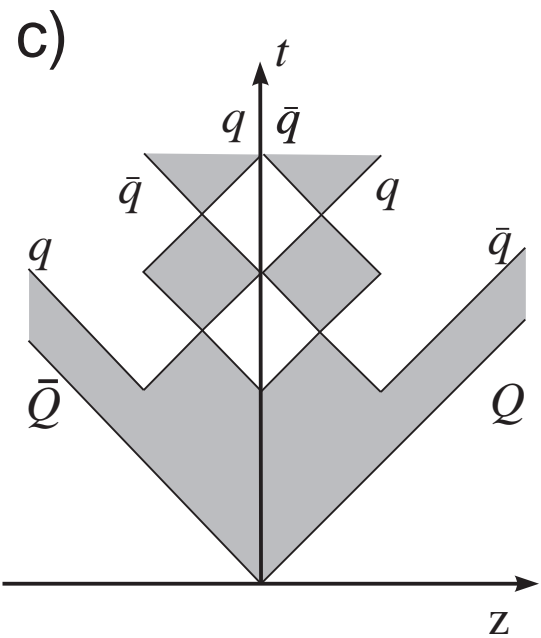
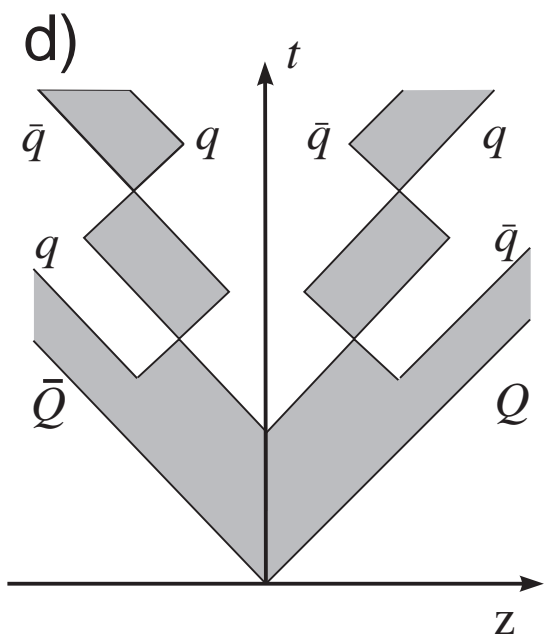
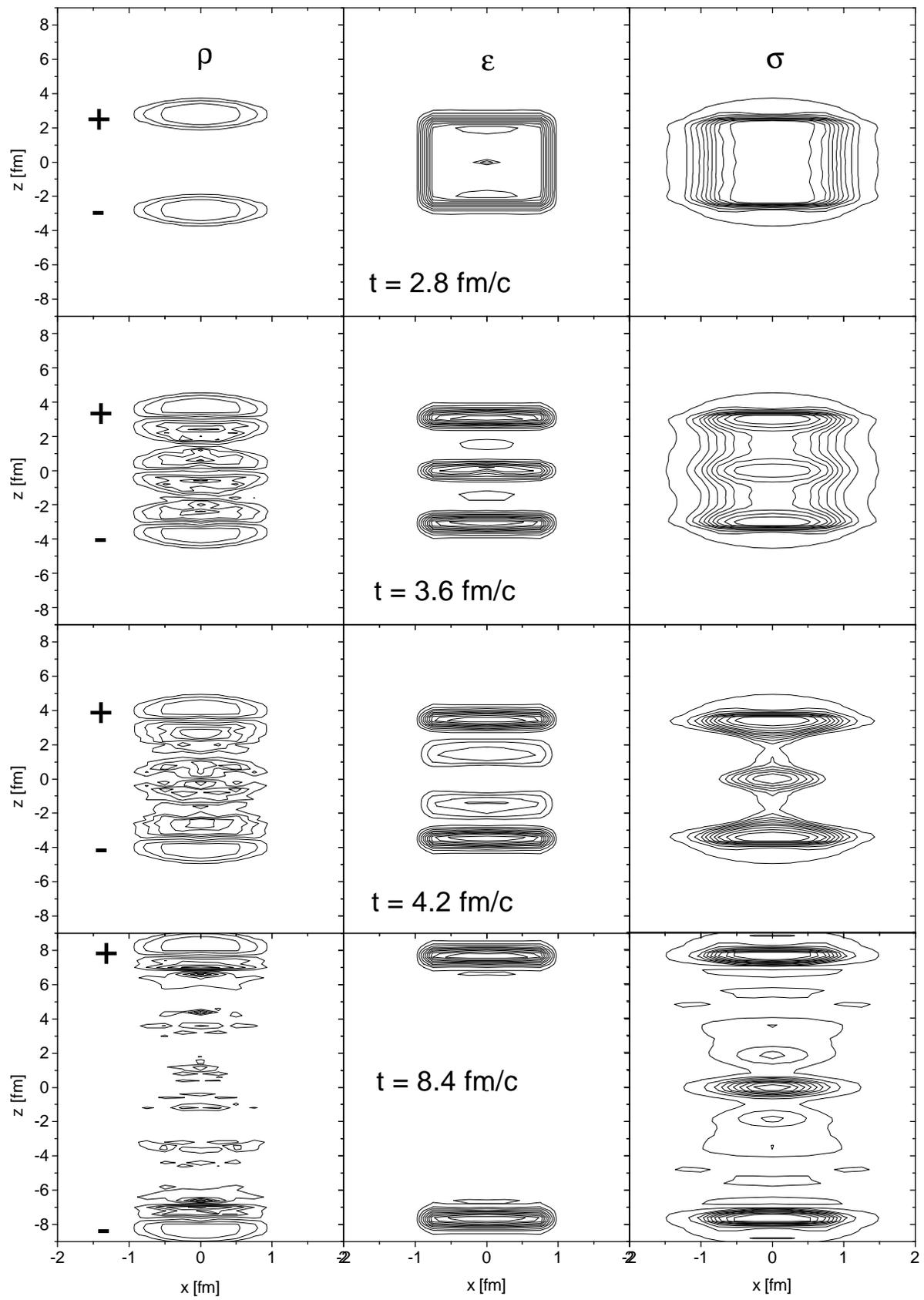
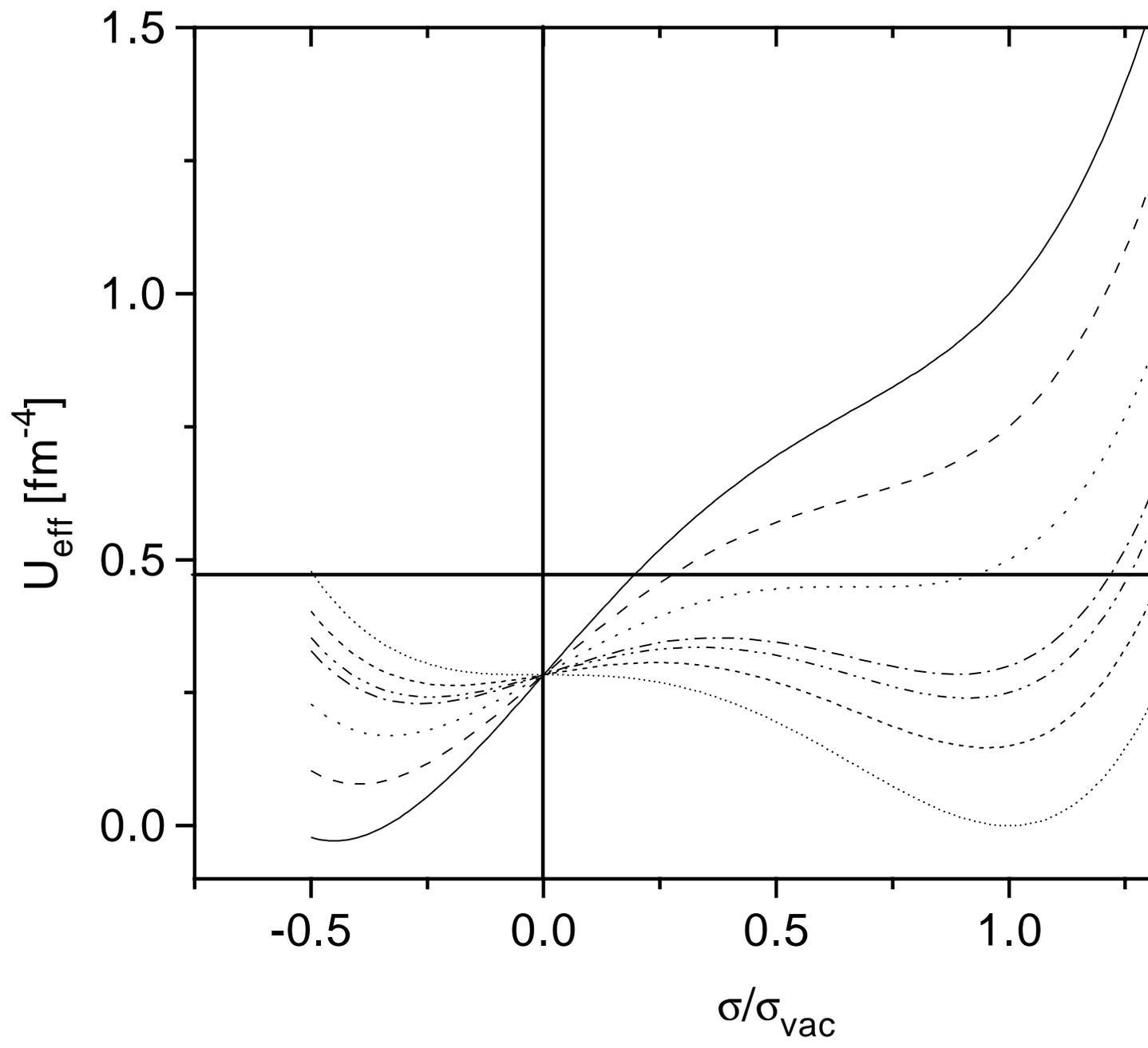
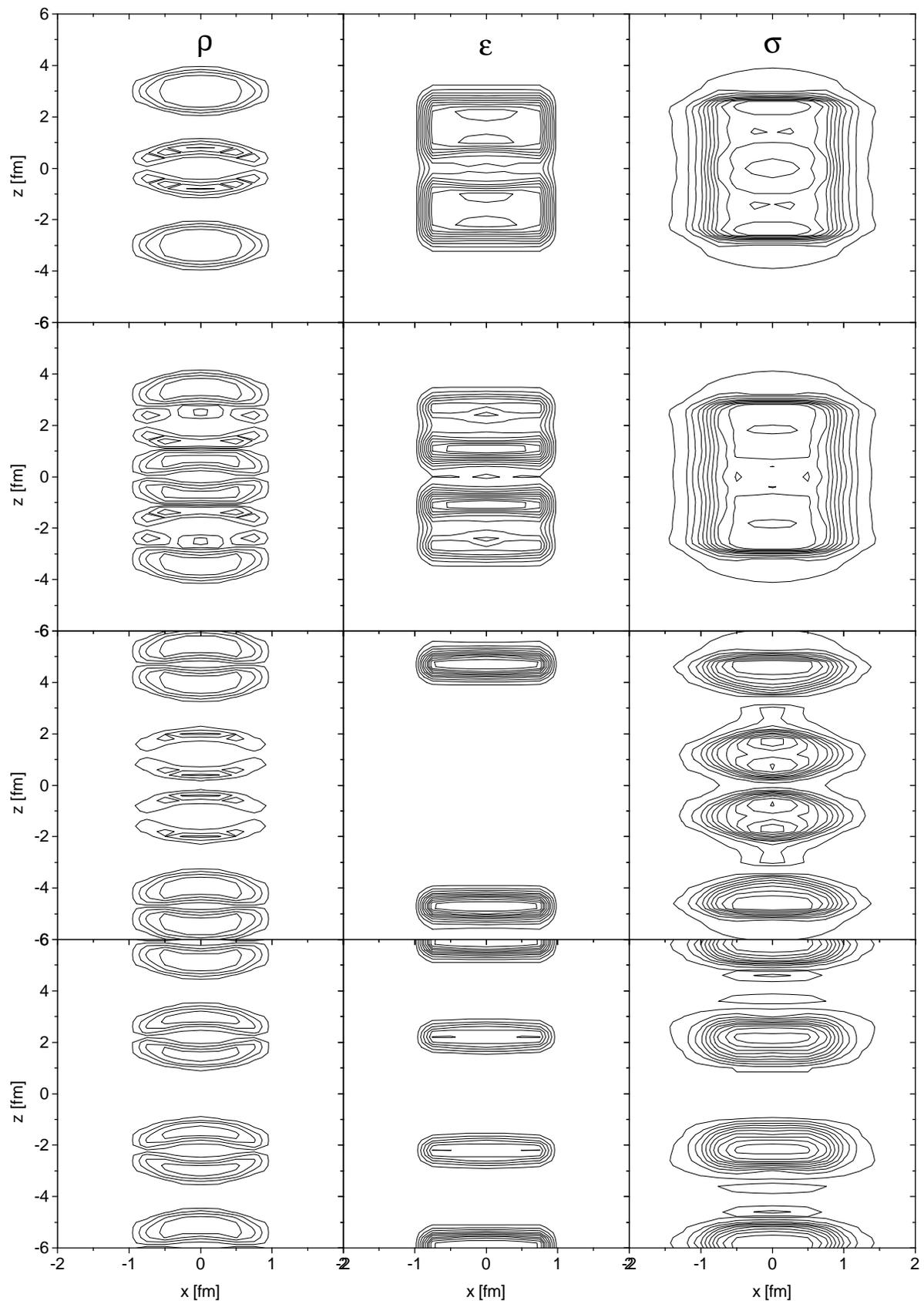
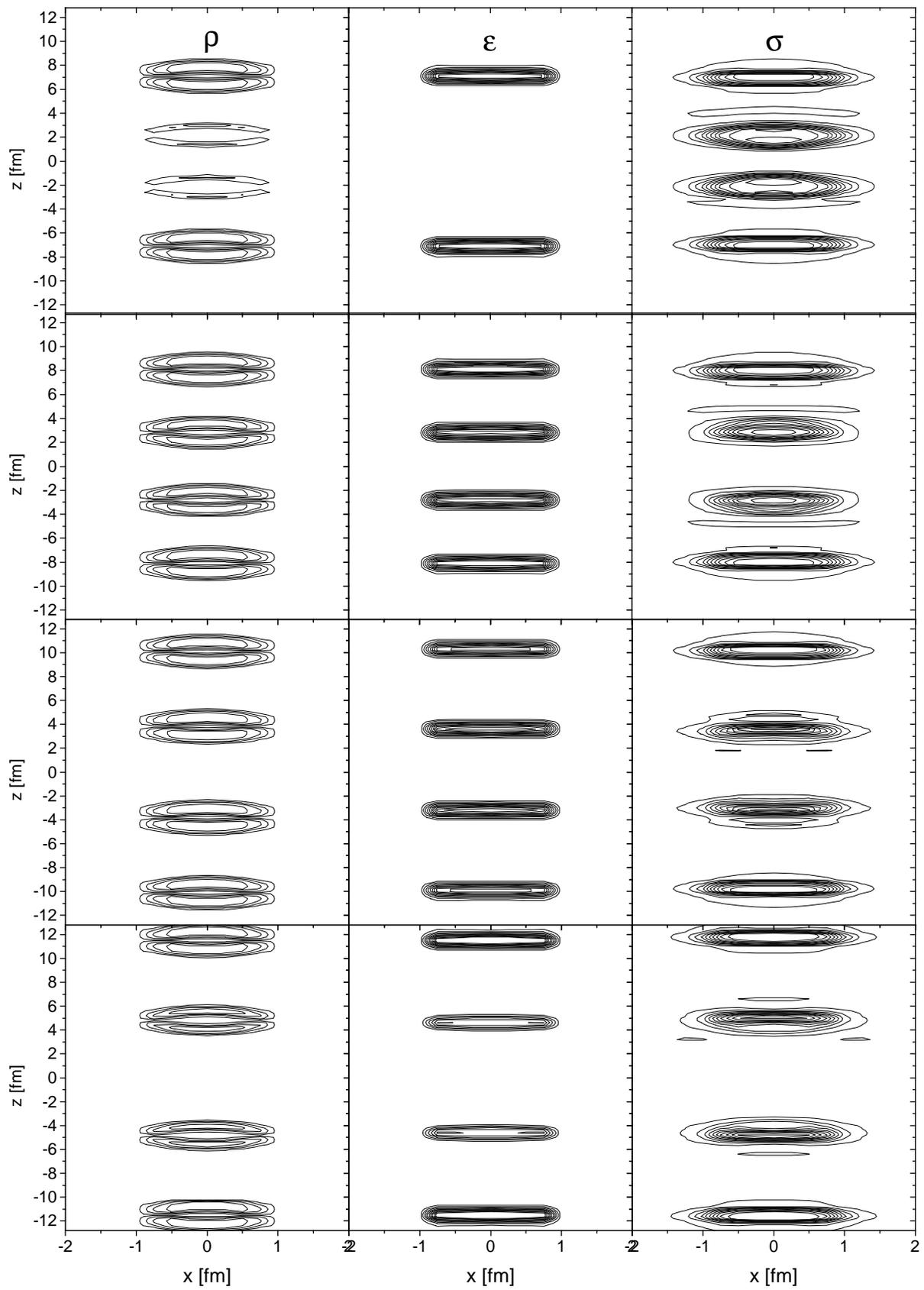
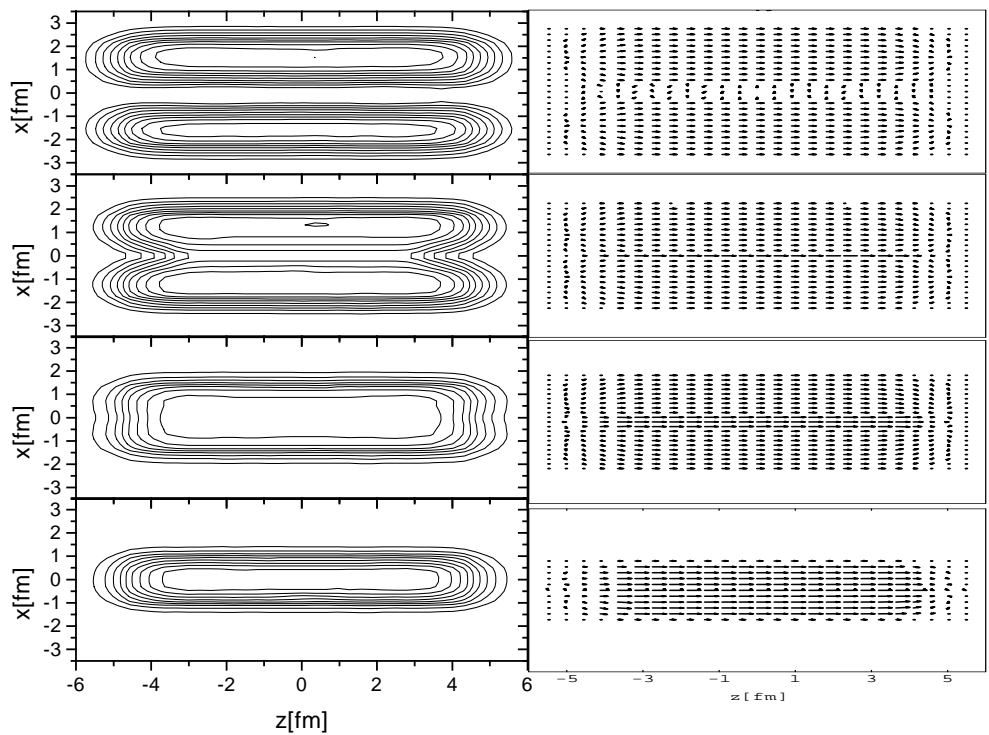
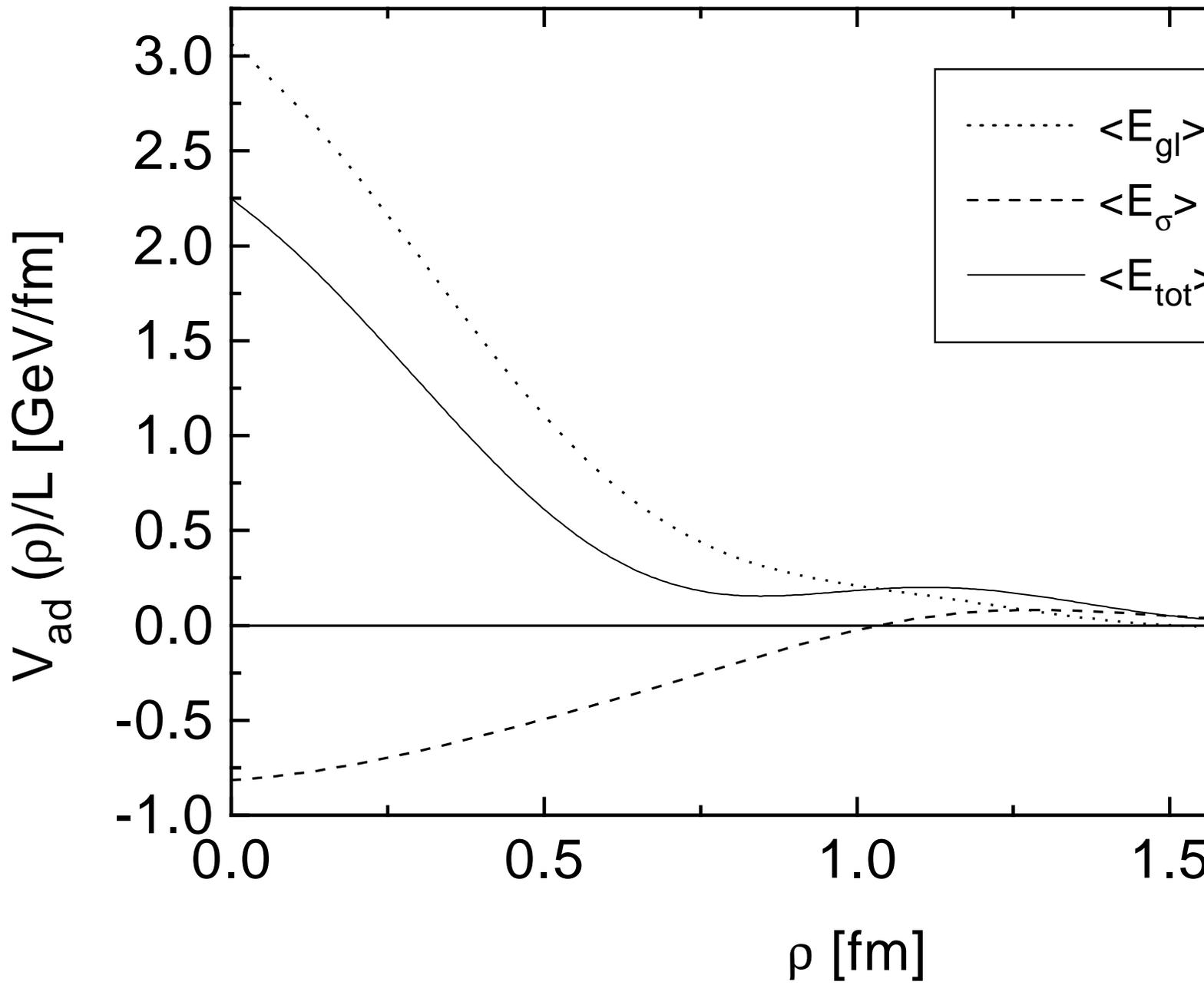
\end{document}